\documentclass[aps,pra,twocolumn,superscriptaddress,showpacs,reprint,groupedaddress]{revtex4-1}
\usepackage{graphicx}
\usepackage{amsfonts}
\usepackage{amssymb}
\usepackage{array}
\usepackage{amsmath}
\usepackage{color}
\usepackage{verbatim}
\usepackage{slashed}
\usepackage{subfigure}
\usepackage{xcolor}
\usepackage{bbold}
\usepackage{lipsum}
\usepackage{mathtools}
\usepackage{bbm}
\DeclarePairedDelimiter\ceil{\lceil}{\rceil}

\makeatletter
\renewcommand{\maketag@@@}[1]{\hbox{\m@th\normalsize\normalfont#1}}%
\makeatother

\begin{document}

\title{
Relation between the Greenberger-Horne-Zeilinger--entanglement cost of preparing a multipartite pure state and its quantum discord}

\author{Seungho Yang}
\affiliation{Center for Macroscopic Quantum Control,
Department of Physics and Astronomy,
Seoul National University, Seoul, 151-742, Korea}
\author{Hyunseok Jeong}
\affiliation{Center for Macroscopic Quantum Control,
Department of Physics and Astronomy,
Seoul National University, Seoul, 151-742, Korea}
\date{\today}

\begin{abstract}
We investigate how much amount of Greenberger-Horne-Zeilinger (GHZ) entanglement is required in order to prepare a given multipartite state by local operations and classical communication (LOCC).We present a LOCC procedure that asymptotically converts GHZ states into an arbitrary multipartite pure state, whose conversion rate is given by the multipartite discord of the state. This reveals that the GHZ-entanglement cost of preparing a pure state is not higher than the multipartite discord of the state. It also provides an operational interpretation of multipartite discord for pure states, namely, the consumption rate of GHZ entanglement in the devised procedure.
\end{abstract}

\maketitle

\section{Introduction} Quantum entanglement lies at the heart of notable features of quantum physics and the power of quantum information processing~\cite{Jozsa03,Bennet92,Bennet93,Giovannetti04}. Its quantification is an issue of great importance. A possible approach is based on the idea that entanglement is a resource shared between distant parties that cannot be generated by local operations and classical communication (LOCC).
More specifically, a scale of entanglement can be created by considering asymptotic conversion (i.e., conversion of $m$ copies of a state into $n$ copies of another state for large $n$ and $m$) under LOCC. Entanglement is then quantified as the optimal rate of the asymptotic conversion, $m/n$, from a resource state to the given state (or vice versa).

Bennett~\emph{et al}. showed that, in the bipartite case, the asymptotic conversion between any pure states is reversible~\cite{Bennet96-1}, so when quantifying entanglement of bipartite systems, any choice of a pure entangled state as a resource state results in an equivalent quantification. The singlet state is a natural choice for a resource state, and the corresponding conversion rate is called the entanglement cost (and it is called the distillable entanglement for the reverse direction)~\cite{Bennet96-1, Bennet96-2, Horodecki09,Plenio07,Horodecki98,Hayden01}. The entanglement cost of preparing a bipartite pure state has been shown to equal the entropy of entanglement or, equivalently, the von Neumann entropy of one of the subsystems~\cite{Bennet96-1}.

 Entanglement of multipartite states can be quantified in the same way, but different choices of the resource state may give rise to independent quantifications. For example, there exist asymptotically inequivalent multipartite states such as the Greenberger-Horne-Zeilinger (GHZ) state and the W state.
Hence, one may choose a resource state and correspondingly define a multipartite entanglement cost as the optimal conversion rate from the resource state to a given state~\cite{Plenio07,A}.
However, even achievable rates (i.e., upper bounds for the optimal rate) are not known for any resource state, except in the case where the singlets shared among the multiple parties are used as resource~\cite{Galvao00}. This can be attributed to the difficulty of analyzing general LOCC~\cite{Chitambar11}. The LOCC conversion between multipartite states has been little studied especially in the asymptotic limit, while there have been studies on the stochastic LOCC conversion~\cite{Yu14,Vrana14} and the conversion under asymptotically non-entangling operations~\cite{Brando08,Brando14} in the asymptotic limit.

 Here, we consider the GHZ entanglement cost of preparing a quantum state of an arbitrary number of parties, where the $k$-partite GHZ state is defined as
    \begin{equation*}
    \vert \text{GHZ}\rangle=\frac{1}{\sqrt{2}}(|0\rangle^{\otimes k}+|1\rangle^{\otimes k}),
    \end{equation*}
    with $|0\rangle$ and $|1\rangle$ denoting orthogonal basis states for local subsystems.
We present a LOCC procedure, $\mathcal{L}_{D}$, for the asymptotic conversion from GHZ states to a multipartite pure state $\psi$, namely,
    \begin{equation*}\label{asymptotic creation}
    GHZ^{\otimes n R_{D}}\xrightarrow{  \mathcal{L}_{D} } \psi^{\otimes n}
    \end{equation*}
 for sufficiently large $n$. We find that the conversion rate of this procedure, $R_{D}$, is given by the multipartite discord of state $\psi$, captured by the relative entropy of discord. This implies that the optimal rate is upper bounded by the relative entropy of discord. Our study further provides an operational interpretation of multipartite discord for pure states, namely, the consumption rate of GHZ entanglement in the devised procedure. In the multipartite setting, quantum discord of pure states is distinct from entanglement (although they are equal for bipartite pure states), but its roles and meanings are not well understood compared to those of bipartite discord in several information tasks~\cite{Gu12,Cavalcanti11,Madhok11,Piani11,Datta08,Chuan12}.

 \section{Definitions} We consider a multipartite system $\mathbf{P}=\{P_{1},\dotsc,P_{k}\}$, consisting of $k$ subsystems of an arbitrary dimension. A $k$-partite quantum state is called fully separable if it can be written in a form of $\sum_{j} p_{j} \rho_{1,j}\otimes \dotsb \otimes \rho_{k,j}$. We denote the set of separable states by $\mathcal{S}$. We also denote an arbitrary orthonormal basis for the $j$-th subsystem by $\{\vert x_{j}\rangle\}$, where $x_{j}$'s are integers from $0$ to $\dim(P_{j})-1$. We can then construct a separable basis for the whole system as $\{\vert x_{1},\dotsc,x_{k}\rangle=\vert x_{1}\rangle \otimes \dotsb \otimes \vert x_{k}\rangle\}$. The basis states for the whole system contain no nonclassical correlation, so we call their classical mixtures classically correlated states. To clarify, a $k$-partite quantum state $\sigma$ is said to be classically correlated if it can be written as $\sigma=\sum_ { x_{1},\dotsc,x_{k}}p( x_{1},\dotsc,x_{k})\vert  x_{1},\dotsc,x_{k} \rangle\langle  x_{1},\dotsc,x_{k} \vert$. We denote  the set of classically correlated states by $\mathcal{C}$.

 Apart from the quantification of entanglement based on the LOCC conversion, there is an axiomatic approach to measuring entanglement. In this approach, an entanglement measure is given as a function of the density matrix that satisfies some desirable properties. One such measure is the relative entropy of entanglement defined as~\cite{Vedral98}
 \begin{equation}\label{entanglement}
 \mathcal{E}_{R}(\rho)=\min_{\sigma\in  {\cal S}}S(\rho \vert\vert\sigma)
 \end{equation}
where $S(\rho\vert\vert\sigma)=-\text{tr}(\rho\log_{2}\sigma)-S(\rho)$ is the relative entropy of $\rho$ to $\sigma$ and $S(\rho)=-\text{tr}(\rho\log_{2}\rho)$ is the von Neumann entropy of $\rho$. When dealing with many copies of a state, it is desirable to consider the regularized version of $\mathcal{E}_{R}$ defined as~\cite{Horodecki00}
  \begin{equation*}
  \mathcal{E}_{R}^{\infty}(\rho)=\lim_{t\rightarrow \infty}\frac{\mathcal{E}_{R}(\rho^{\otimes t})}{t}.
  \end{equation*}

Using the set of classically correlated states, $\mathcal{C}$, the relative entropy of discord~\cite{Modi10,SaiToh08,Groisman07} and its regularized versions are defined as
     \begin{equation}
    \label{discord}
    \mathcal{D}_{R}(\rho)=\min_{\sigma\in  \mathcal{C}}S(\rho \vert\vert\sigma), ~~~ \mathcal{D}^{\infty}_{R}(\rho)=\lim_{t\rightarrow \infty}\frac{\mathcal{D}_{R}(\rho^{\otimes t})}{t},
    \end{equation}
    respectively. From the definitions, it is evident that $\mathcal{D}^{\infty}_{R}\leqslant \mathcal{D}_{R}$ and $\mathcal{E}^{\infty}_{R}\leqslant \mathcal{E}_{R}$. As already mentioned, $\mathcal{D}_{R}=\mathcal{E}_{R}$ for any bipartite pure state, but they are generally not equal for $k$-partite pure states with $k>2$.
It has been shown that one can always find a separable basis $\{\vert x_{1},\dotsc,x_{k}\rangle\}$ such that the complete set of projectors $\{\Pi_{j}\}=\{\vert x_{1},\dotsc,x_{k}\rangle\langle x_{1},\dotsc,x_{k}\vert \}$ satisfies $\mathcal{D}_{R}(\rho)=S(\rho\vert\vert \sum_{j} \Pi_{j}\rho\Pi_{j})$~\cite{Modi10}.
 We then obtain
    \begin{equation}
    \label{discord_form}
    \mathcal{D}_{R}(\psi)=\min_{\{\Pi_{j}\}} \left[-\sum\text{tr}(\Pi_{j}\psi)\log_{2} \text{tr}(\Pi_{j}\psi)\right]
    \end{equation}
 for a pure state $\psi$.

 To address the GHZ entanglement cost, we define the asymptotic preparation as follows. We say that a LOCC procedure, denoted by $\mathcal{L}$, asymptotically prepare $\psi$ from GHZ states at rate $R$ if $F\textbf{(}(\psi^{\otimes n}, \mathcal{L}(GHZ^{\otimes m})\textbf{)}\rightarrow 1$ and $m/n\rightarrow R$ as $n\rightarrow \infty$, where $F(\rho , \sigma)=\text{tr}\sqrt{\rho ^{\frac{1}{2}}\sigma\rho ^{\frac{1}{2}}}$ is the fidelity between two quantum states $\rho$ and $\sigma$~\cite{Uhlman76}.

 \section{Main result}
  We now provide our main result as follows. First, we introduce a pure state $\Psi$ that approximates $\psi^{\otimes n}$. It is based on the asymptotic equipartition property (AEP)~\cite{Cover91} which will be briefly explained. Next, we present a LOCC procedure that prepares $\Psi$ from a certain number of copies of a GHZ state, say $m$ copies. Finally, we prove that $F(\Psi, \psi^{\otimes n})\rightarrow 1$ and $m/n\rightarrow \mathcal{D}^{\infty}_{R}(\psi)$ for $n\rightarrow \infty$, so they collectively verify that the LOCC procedure asymptotically prepare $\psi$ at rate $\mathcal{D}^\infty_{R}(\psi)$.
 \subsection{Introduction of an approximate state}
Here we introduce a pure state $\Psi$ that approximates $\psi^{\otimes n}$ (i.e., $F(\psi^{\otimes n}, \Psi)\rightarrow 1$ for $n\rightarrow\infty$). For simplicity, we only consider tripartite systems, but the generalization to any $k$-partite system is straightforward.

We begin with summarizing the AEP \cite{Cover91}. Consider independent and identically distributed random variables $X^{(1)},\dotsc, X^{(l)}$. Each of the variables has the same probability distribution $p(x)$ and the Shannon entropy $H$.
 We define a typical set $\mathcal{A}_{\epsilon}$ to be a set of sequences $\{x^{l}=(x^{(1)},\dotsc,x^{(l)})\}$ that satisfy $2^{-l(H+\epsilon)}\leqslant p(x^{l})\leqslant 2^{-l(H-\epsilon)}$. Then, the AEP states that the typical sequences $\{x^{l}\}$ contain most of the probability, and the size of the typical set, $\vert\mathcal{A}_{\epsilon}\vert$, is about $2^{H}$. The AEP is summarized as follows.
 For any $\epsilon>0$,
     \begin{equation}
     \begin{split}
     \text{Pr}\left[x^{l}\in \mathcal{A}_{\epsilon}\right]&>1-\epsilon, \\
     \vert\mathcal{A}_{\epsilon}\vert~~~ &> (1-\epsilon)2^{l(H-\epsilon)}, \\
     \vert\mathcal{A}_{\epsilon}\vert~~~ &\leqslant 2^{l(H+\epsilon)}
     \end{split}
     \end{equation}
 for sufficiently large $l$.

Let us assume that we want to asymptotically prepare  a tripartite state $\psi$, where the dimensions of the three subsystems are $\bar l$, $\bar m$ and $\bar n$. We then consider the state $\phi=\psi^{\otimes t}$ prepared in a tripartite system $\mathbf{P}=\{P_{1}, P_{2}, P_{3}\}$ where
$\dim(P_{1})=({\bar l})^{t}$, $\dim(P_{2})=({\bar m})^{t}$ and $\dim(P_{3})=({\bar n})^{t}$. This state can be written in a separable basis $\{\vert x_{1},\dotsc,x_{k}\rangle\}$ for system $\mathbf{P}$ as
    \begin{equation}\label{local rep}
    \vert\phi\rangle=\vert \psi \rangle^{\otimes t} =\sum_{x_{1},x_{2},x_{3}}C(x_{1},x_{2},x_{3}) \vert x_{1},x_{2},x_{3}\rangle
    \end{equation}
 with some coefficients $C(x_{1},x_{2},x_{3})$. We note that $\{\vert x_{j}\rangle\}$'s can be any orthonormal bases for each subsystem, and we do not choose any particular separable basis at this point. The coefficients $\vert C(x_{1},x_{2},x_{3})\vert^{2}$ can be considered a joint probability distribution of random variables $X_{1}$, $X_{2}$, and $X_{3}$, so we set $p(x_{1},x_{2},x_{3})=\vert C(x_{1},x_{2},x_{3})\vert^{2}$ and
  \begin{equation}\label{entropy}
  H=-\sum\vert C(x_{1},x_{2},x_{3})\vert^{2}\log_{2} \vert C(x_{1},x_{2},x_{3})\vert^{2}.
  \end{equation}
 We may consider independent and identically distributed random variables $X^{(1)}_{j},\dotsc,X^{(l)}_{j}$ for $j=$1, 2, and 3. A typical set is correspondingly defined as
 \begin{equation}\label{typical}
 \mathcal{A}_{\epsilon}=\left\{(x_{1}^{l},x_{2}^{l},x_{3}^{l}): \left\vert \log_{2}  p(x_{1}^{l},x_{2}^{l},x_{3}^{l})-H \right\vert \leqslant \epsilon\right\}.
 \end{equation}

  Then, $n=t\cdot l$ copies of $\psi$, which is prepared in a system $\mathbf{Q}=\{Q_{1}, Q_{2}, Q_{3}\} = \{P_{1}^{\otimes l}, P_{2}^{\otimes l}, P_{3}^{\otimes l}\}$, is represented using the typical set $\mathcal{A}_{\epsilon}$ and its complement $\mathcal{A}_{\epsilon}^{c}$ as
     \begin{equation} \label{orginiral}
     \begin{split}
     \vert\psi\rangle^{\otimes n}_{\mathbf{Q}}=\sum_{(x_{1}^{l},x_{2}^{l},x_{3}^{l}) \in \mathcal{A}_{\epsilon}} C(x_{1}^{l},x_{2}^{l},x_{3}^{l}) \vert x_{1}^{l},x_{2}^{l},x_{3}^{l}\rangle \\+ \sum_{(x_{1}^{l},x_{2}^{l},x_{3}^{l}) \in \mathcal{A}_{\epsilon}^{c}} C(x_{1}^{l},x_{2}^{l},x_{3}^{l}) \vert  x_{1}^{l},x_{2}^{l},x_{3}^{l}\rangle.
     \end{split}
     \end{equation}

  The approximated state $\Psi$ is defined to consist of only the terms corresponding to the typical sequences, so
     \begin{equation} \label{approximate}
     \vert\Psi\rangle_{\mathbf{Q}} =N_{\epsilon}^{-\frac{1}{2}}\sum_{(x_{1}^{l},x_{2}^{l},x_{3}^{l}) \in \mathcal{A}_{\epsilon}} C(x_{1}^{l},x_{2}^{l},x_{3}^{l})\vert x_{1}^{l},x_{2}^{l},x_{3}^{l}\rangle
     \end{equation}
  where $N_{\epsilon}=\sum_{(x_{1}^{l},x_{2}^{l},x_{3}^{l}) \in \mathcal{A}_{\epsilon}} \vert C(x_{1}^{l},x_{2}^{l},x_{3}^{l})\vert ^{2}$ is a normalization factor. As the typical set contains most of the probability, we see that the fidelity between $\Psi$ and $\psi^{\otimes n}$ approaches unity as $n\rightarrow \infty$.
  In addition, it follows from the AEP that the number of terms in the expansion of $\Psi$, $\vert\mathcal{A}_{\epsilon}\vert$, is approximately $2^{lH}$.

At this point, we note a previous method for bipartite asymptotic preparation~\cite{Bennet96-1}
in relation to our approach. In Ref.~\cite{Bennet96-1}, one party, say the first party, prepares $\psi^{\otimes n}_{Q_{1}Q_{1}'}$ at their site and send the compressed $Q_{1}'$-part (the compression uses the AEP) to the second party using the quantum teleporation protocol. At the end of the procedure, the two parties share an approximated state of the same form as $\Psi$. This method using the teleportation protocol can be applied only to multipartite states that have Schmidt decompositions such as $\alpha \vert 000\rangle +\beta |111\rangle$~\cite{Bennet00}. On the other hand, our LOCC procedure directly converts singlets into the state $\Psi$. It enables us to consider the approximated state $\Psi$ in multipartite setting, which can actually be created by LOCC as shown in the next subsection.

 \subsection{Preparation of the approximate state}
We now present the LOCC procedure $\mathcal{L}_{D}$ that converts $\log_{2} \vert \mathcal{A}_{\epsilon}\vert$ copies of a GHZ state into $\Psi$, so $\mathcal{L}_{D}(GHZ^{\otimes \log_{2} \vert \mathcal{A}_{\epsilon}\vert})=\Psi$. We also only consider tripartite systems, and the generalization to any $k$-partite system is straightforward. For convenience of mathematical descriptions, we change the notation as follows. For $x_{1}^{l}$, $x_{2}^{l}$, and $x_{3}^{l}$ such that $(x_{1}^{l},x_{2}^{l},x_{3}^{l}) \in \mathcal{A}_{\epsilon}$, consider the following sets: $\{(x_{1}^{l},x_{2}^{l},x_{3}^{l})\}$, $ \{x_{1}^{l}\}$, $\{x_{2}^{l}\}$, and $\{x_{3}^{l}\}$, whose sizes are $\vert \mathcal{A}_{\epsilon}\vert$, $\alpha$, $\beta$ and $\gamma$, respectively. We replace $(x_{1}^{l}, x_{2}^{l}, x_{3}^{l})$ with $y$ where $y\in \{0, 1,\dotsc , \vert \mathcal{A}_{\epsilon}\vert-1\}$. Similarly, we do the same for other sets as
 \begin{equation}
 \begin{split}
 \{x_{1}^{l}\}&\rightarrow \{f: f\in 0,\dotsc,\alpha-1\},\\
 \{x_{2}^{l}\}&\rightarrow \{g: g\in 0,\dotsc, \beta-1\},\\
 \{x_{3}^{l}\}&\rightarrow \{h: h\in 0,\dotsc, \gamma-1\}.
 \end{split}
 \end{equation}
  Because $f$, $g$, and $h$ are completely determined by $y$, we denote them by $f(y)$, $g(y)$, and $h(y)$. We can then rewrite the state $\Psi$ of the system $\mathbf{Q}$ as
     \begin{equation} \label{output}
     \vert\Psi\rangle_{\mathbf{Q}}=N_{\epsilon}^{-\frac{1}{2}}\sum_{y=0}^{\vert \mathcal{A}_{\epsilon}\vert-1} C(y)\vert f(y), g(y), h(y)\rangle_{\mathbf{Q}}.
     \end{equation}

The state $\Psi$ can be obtained from $m=\ceil*{ \log_{2}\vert\mathcal{A}_{\epsilon}\vert}$ copies of a GHZ state ($\ceil*{x}$ is the smallest integer not less than $x$) by following four steps of LOCC. Assume that $m$ copies of a GHZ state are prepared in an ancillary system $\mathbf{Q}'=\{Q_{1}', Q_{2}', Q_{3}'\}$, where $\dim(Q_{1})=\dim(Q_{2})=\dim(Q_{3})=\vert \mathcal{A}_{\epsilon}\vert$. We can rewrite $GHZ^{\otimes m}$ as
    \begin{equation*}
    \begin{split}
    \vert \text{GHZ}\rangle_{\mathbf{Q}'}^{\otimes m}&= 2^{-\frac{m}{2}}\sum_{z_{j}=0,1}\vert z_{1}\dotsb z_{m}\rangle \vert z_{1}\dotsb z_{m}\rangle\vert z_{1}\dotsb z_{m}\rangle \\&= 2^{-\frac{m}{2}}\sum_{y=0}^{2^{m}-1}\vert y, y, y \rangle_{\mathbf{Q}'},
    \end{split}
    \end{equation*}
where $y$ is the decimal representation of binary strings $z_{1}\dotsb z_{m}$. Because $\vert\mathcal{A}_{\epsilon}\vert\leqslant 2^{m}$, we may discard the terms other than those with $0\leqslant y \leqslant \vert\mathcal{A}_{\epsilon}\vert-1$ (by a simple local operation), so we have
   \begin{equation*}
   \vert \mathcal{A}_{\epsilon}\vert^{-\frac{1}{2}}\sum_{y=0}^{\vert \mathcal{A}_{\epsilon}\vert-1}\vert y, y, y\rangle_{\mathbf{Q}'} .
    \end{equation*}

 The first step is to change the coefficients from $ 2^{-\frac{m}{2}}$ to $N_{\epsilon}^{-1/2}C(y)$ as
 \begin{small}
 \begin{equation*}
  \vert \mathcal{A}_{\epsilon}\vert^{-\frac{1}{2}}\sum_{y=0}^{\vert \mathcal{A}_{\epsilon}\vert-1}\vert y, y, y\rangle_{\mathbf{Q}'}  \xrightarrow{{\rm Step} 1} N_{\epsilon}^{-\frac{1}{2}} \sum_{y=0}^{\vert \mathcal{A}_{\epsilon}\vert-1} C(y)\vert y, y, y \rangle_{\mathbf{Q}'}.
  \end{equation*}
 \end{small}

  This can be done by local operations of any party. Consider a local measurement described by the measurement operators
 \begin{small}
 \begin{equation}
 \left\{M_{j}=N_{\epsilon}^{-\frac{1}{2}}\sum_{y=0}^{\vert \mathcal{A}_{\epsilon}\vert-1}D(y\oplus j)\vert y \rangle \langle y \vert, ~0 \leqslant j \leqslant \vert \mathcal{A}_{\epsilon}\vert-1 \right\}
  \end{equation}
 \end{small}
  where $\oplus$ denotes addition modulo $\vert \mathcal{A}_{\epsilon}\vert$. It is easy to check that they satisfy the completeness relation $\sum_{j} M{^\dagger}_{j}M_{j}=I$. After the measurement, if the outcome of the measurement is $j$, each of the parties applies a unitary operation $U$: $\vert y\rangle \rightarrow \vert y\oplus j\rangle$ to complete the first step. The second step is to create the state $\vert f(y), h(y), g(y)\rangle_{\mathbf{Q}}$ in the system $\mathbf{Q}$, which is initially prepared in $\vert 000\rangle_{\mathbf Q}$.
    \begin{equation*}
    \begin{split}
    \xrightarrow{{\rm Step} 2} N_{\epsilon}^{-\frac{1}{2}}\sum_{y=0}^{\vert \mathcal{A}_{\epsilon}\vert-1} C(y)\vert y, y, y\rangle_{\mathbf{Q}'} \otimes  \vert f(y), h(y), g(y)\rangle_{\mathbf{Q}}.
    \end{split}
    \end{equation*}
 This is achieved by a local unitary operation on $Q_{1}Q_{1}'$ that transforms $\vert y \rangle_{Q_{1}'} \otimes\vert 0\rangle_{Q_{1}}$ into $\vert y\rangle_{Q_{1}'}\otimes \vert f(y)\rangle_{Q_{1}}$, and similar local unitary operations on $Q_{2}Q_{2}'$ and $Q_{3}Q_{3}'$.

  The third step is to disentangle $Q_{2}'$ and $Q_{3}'$ as
    \begin{equation}\label{disentangle}
    \xrightarrow{{\rm Step} 3} N_{\epsilon}^{-\frac{1}{2}}\sum_{y=0}^{\vert \mathcal{A}_{\epsilon}\vert-1} C(y)\vert y\rangle_{Q_{1}'} \otimes  \vert f(y), g(y), h(y)\rangle_{\mathbf{Q}}.
    \end{equation}
In oder to perform this step, the second and third parties perform local measurements on their systems using the measurement operators
 \begin{equation*}
 \{M_{j}=\vert \mathcal{A}_{\epsilon}\vert^{-1} J \vert j \rangle \langle j \vert J^{\dagger}\},
 \end{equation*}
 where $J$ is a complex Hadamard operation, defined as $\langle y\vert J \vert y' \rangle =\exp [2\pi i\cdot yy'/\vert \mathcal{A}_{\epsilon}\vert]$. Depending on the measurement outcomes, the first party can perform a phase-shifting operation to complete the third step.

 The final step is to disentangle the ancillary system $Q_{1}'$ from $\mathbf{Q}$ to obtain $\Psi_{Q}$ as
  \begin{equation}\label{to}
   \xrightarrow{{\rm Step} 4}N_{\epsilon}^{-\frac{1}{2}}\sum_{y=0}^{\vert \mathcal{A}_{\epsilon}\vert-1} C(y)\vert f(y), g(y), h(y)\rangle_{\mathbf{Q}}=\vert\Psi\rangle_{\mathbf{Q}}.
   \end{equation}

  To address the procedure for this step, we introduce a variable $K(y)$, which is defined as
 \begin{equation}\label{q}
 K(y)=\beta\gamma f(y)+\gamma g(y)+h(y).
 \end{equation}
Then, we can consider a linear isometry transformation $V$ on the system $Q_{1}'$ which transforms $\vert y \rangle$ into $\vert K(y) \rangle$. 
This requires additional ancillary qubits in $Q_{1}'$ so that the dimension of $Q_{1}'$ equals $K(\vert \mathcal{A}_{\epsilon}\vert-1)+1$.
 Applying $V$ to the state in Eq.~\eqref{disentangle} gives
     \begin{equation*}
    \sum _{y=0}^{\vert \mathcal{A}_{\epsilon}\vert-1} C(y) \vert K(y) \rangle_{Q_{1}'} \otimes  \vert f(y), g(y), h(y)\rangle_{\mathbf{Q}}.
     \end{equation*}
 Next, the first party performs a measurement on the ancillary system $Q_{1}'$ using the operators
 \begin{small}
 \begin{equation*}
 \left\{M_{j}: M_{j}=\frac{\vert0\rangle\langle j\vert J^{\dagger}}{[K(\vert \mathcal{A}_{\epsilon}\vert-1)+1]^{\frac{1}{2}}} ,~0\leqslant j \leqslant K(\vert \mathcal{A}_{\epsilon}\vert-1)\right\},
 \end{equation*}
 \end{small}
 where $\tilde{J}$ is a complex Hadamard operation defined as
 \begin{equation*}
  \langle y\vert \tilde{J} \vert y' \rangle =\exp \left[2\pi i ~ y y'~[K(\vert \mathcal{A}_{\epsilon}\vert-1)+1]^{-1}\right]
 \end{equation*}
  for $0\leqslant y,~y' \leqslant K(\vert \mathcal{A}_{\epsilon}\vert-1)+1$. The completeness relation, $\sum _{j} M_{j}^{\dagger} M_{j}=I$, can be checked from the orthogonality of the Hadamard operation, $\tilde{J}\tilde{J}^{\dagger}=[K(\vert \mathcal{A}_{\epsilon}\vert-1)+1]~ I$. If the measurement outcome is $j$, the resulting state is
     \begin{equation*}
     \begin{split}
     &\vert0\rangle_{Q_{1}'}\sum _{y=0}^{\vert \mathcal{A}_{\epsilon}\vert-1} C(y) \langle j \vert J^{\dagger} \vert K(y) \rangle \otimes  \vert f(y), g(y), h(y)\rangle_{\mathbf{Q}}\\
     =&\vert0\rangle_{Q_{1}'}\sum _{y=0}^{\vert \mathcal{A}_{\epsilon}\vert-1}C(y)e^{-2\pi i ~ j K(y)} \otimes  \vert f(y), g(y), h(y)\rangle_{\mathbf{Q}}.
     \end{split}
     \end{equation*}

  Finally, local phase-shifting operations can remove the phase $\exp[-2\pi i ~ j K(y)]$ by using Eq.~\eqref{q}. Consider a phase-shifting operation by the first party that transforms
  \begin{equation*}
  \vert f(y)\rangle \rightarrow \exp[2\pi i ~ j\beta\gamma f(y)]\vert f(y)\rangle.
  \end{equation*}

  Similarly, consider phase-shifting operations by the second and the third parties that transform
  \begin{equation*}
  \begin{split}
  \vert g(y)\rangle &\rightarrow \exp[2\pi i ~ j\gamma g(y)]\vert g(y)\rangle, \\
  \vert h(y)\rangle &\rightarrow \exp[2\pi i ~ j h(y)]\vert h(y)\rangle.
  \end{split}
  \end{equation*}
  Applying those operations completes the fourth step.


 \subsection{Asymptotic preparation from GHZ states at a rate equal to quantum discord}
We denote by $\mathcal{L}_{D}$ the LOCC procedure described in the previous subsection. We have shown that it prepares $\Psi$ from $GHZ^{\otimes \ceil*{\log_{2} \vert \mathcal{A}_{\epsilon}\vert}}$, i.e., $\mathcal{L}_{D}(GHZ^{\otimes \ceil*{\log_{2}\vert \mathcal{A}_{\epsilon}\vert}})=\Psi$.
 In practice, it can asymptotically prepare $\psi$ from GHZ states at rate $\mathcal{D}^{\infty}_{R}$. This can be explained as follows.
First, the AEP implies that the fidelity $F(\Psi,\psi^{\otimes n})$ approaches unity as $n\rightarrow \infty$, so $\mathcal{L}_{D}$ asymptotically prepares $\psi^{\otimes n}$. Second, the conversion rate, which is given by $n^{-1}\ceil*{\log_{2}\vert \mathcal{A}_{\epsilon}\vert}$, can be reduced down to $\mathcal{D}^{\infty}_{R}(\psi)$. Recall that the typical set in Eq.~\eqref{local rep} depends on the separable basis $\{x_{1}, x_{2}, x_{3}\}$, which we have not specified yet. We now choose the separable basis so that it gives the minimum size of the typical set. Let us denote the minimum size by $\log_{2}\vert\mathcal{A}_{\epsilon}^{*}\vert$ and denote the corresponding Shannon entropy by $H^{*}$. It then follows from the AEP that $\log_{2}\vert\mathcal{A}_{\epsilon}^{*}\vert\approx l H^{*}$. In addition, a comparison between Eqs.~\eqref{discord_form} and \eqref{entropy} leads to $H^{*}= \mathcal{D}_{R}(\psi^{\otimes t})$. Putting these together, we have $n^{-1}\ceil*{\log_{2}\vert \mathcal{A}_{\epsilon}^{*}}\vert =t^{-1}l^{-1}\ceil*{\log_{2}\vert \mathcal{A}_{\epsilon}^{*}\vert}\approx t^{-1} \mathcal{D}_{R}(\psi^{\otimes t})$. Finally, from the definition of $\mathcal{D}^{\infty}_{R}$, the conversion rate is approximately found to be $\mathcal{D}^{\infty}_{R}(\psi)$ for large $t$. We thus reach the following theorem.

  \textbf{Theorem.} The LOCC procedure $\mathcal{L}_{D}$ asymptotically prepares a multipartite pure state $\psi$ from GHZ states at rate $\mathcal{D}_{R}^{\infty}(\psi)$. Namely, for any $\epsilon>0$, $\delta>0$,
    \begin{equation}\label{theorem}
    \begin{split}
    F\textbf{(}\mathcal{L}_{D}(GHZ^{\otimes \ceil*{\log_{2}\vert \mathcal{A}^{*}_{\epsilon}\vert}}), \psi^{\otimes n}\textbf{)}&> 1-\epsilon, \\
    \vert n^{-1}\ceil*{\log_{2}\vert \mathcal{A}^{*}_{\epsilon}\vert}- \mathcal{D}^{\infty}_{R}(\psi)\vert &< \delta.
    \end{split}
    \end{equation}
 for sufficiently large $n$.\\

 ~\emph{proof}---
  The LOCC procedure, $\mathcal{L}_{D}$, works for any separable basis $\{\vert x_{1}, x_{2}, x_{3}\rangle\}$ in Eq.~\eqref{local rep}. In a given separable basis, the Shannon entropy of the random variables $X_{1}$, $X_{2}$, and $X_{3}$ is given as
 \begin{equation*}
 H=-\sum \vert C(x_{1},x_{2},x_{3})\vert^{2}\log_{2} \vert C(x_{1},x_{2},x_{3})\vert^{2},
 \end{equation*}
 where $\vert C(x_{1},x_{2},x_{3})\vert^{2}=\text{tr}(\vert x_{1},x_{2},x_{3}\rangle\langle x_{1},x_{2},x_{3}\vert \psi^{\otimes t})$. 
 We note again that the typical set $\mathcal{A}_{\epsilon}$ depends on the separable basis and so does the Shannon entropy.
 We can make $H$ equal $\mathcal{D}_{R}(\psi^{\otimes t})$ by choosing a suitable separable basis.
 Using the expression of $\mathcal{D}_{R}$ in Eq.~\eqref{discord_form}, we can write
 \footnotesize
 \begin{equation*}
 \begin{split}
 \mathcal{D}_{R}(\psi^{\otimes t})=\min_{\substack{\{\vert x_{1}, x_{2}, x_{3}\rangle\langle  x_{1}, x_{2}, x_{3} \vert \}}}\Big[-\sum \text{tr}(\vert x_{1},x_{2},x_{3}\rangle\langle x_{1},x_{2},x_{3}\vert \\ \times \psi^{\otimes t})\log_{2} \text{tr}(\vert x_{1},x_{2},x_{3}\rangle\langle x_{1},x_{2},x_{3}\vert \psi^{\otimes t})\Big].
 \end{split}
 \end{equation*}
 \normalsize
  Therefore, by choosing a separable basis that attains the minimum in the above equation, we have $H=\mathcal{D}_{R}(\psi^{\otimes t})$. For clarity, we add a superscript * to the corresponding typical set and Shannon entropy as $\mathcal{A}^{*}_{\epsilon}$ and $H^{*}$. Then, it follows from the AEP that, for any $\epsilon' >0$,
     \begin{equation} \label{AEP}
     \begin{split}
       N_{\epsilon'}&> (1-\epsilon'), \\
     (1-\epsilon')2^{l (H^{*}-\epsilon')}< \vert \mathcal{A}^{*}_{\epsilon'}\vert&\leqslant  2^{l (H^{*}+\epsilon')}
     \end{split}
     \end{equation}
  for sufficiently large $l$. The fidelity between $\psi^{\otimes n}$ in Eq.~\eqref{orginiral} and $\Psi=\mathcal{L}_{D}(GHZ^{\otimes \ceil*{\log\vert \mathcal{A}^{*}_{\epsilon'}\vert}})$ in Eq.~\eqref{approximate} is given as
     \begin{equation*}
     F(\Psi, \psi^{\otimes n})=\sum_{(x_{1}^{l},x_{2}^{l},x_{3}^{l}) \in \mathcal{A}_{\epsilon}}\frac{\vert  C(x_{1}^{l},x_{2}^{l},x_{3}^{l})\vert^{2}}{N_{\epsilon'}^{\frac{1}{2}}}=N_{\epsilon'}^{\frac{1}{2}}
     \end{equation*}
  In addition, applying the first inequality in Eq.~\eqref{AEP} gives $N_{\epsilon'}^{\frac{1}{2}}>(1-\epsilon')^{\frac{1}{2}}$. By choosing $\epsilon'$ such that $\epsilon'<1-(1-\epsilon)^2$ for any $\epsilon$, we obtain
  \begin{equation*}
  F\textbf{(}\mathcal{L}_{D}(GHZ^{\otimes \ceil*{\log\vert \mathcal{A}^{*}_{\epsilon'}\vert}}), \psi^{\otimes n}\textbf{)}>1-\epsilon
  \end{equation*}
  for any $\epsilon$.

    As the LOCC procedure asymptotically prepares $\psi^{n}$ from $GHZ^{\ceil*{\log_{2} \vert \mathcal{A}^{*}_{\epsilon}\vert}}$, 
  the conversion rate is given by $n^{-1}\log_{2} \vert \mathcal{A}^{*}\vert$ with $n=t~l$ (the ceiling function $\ceil*{~}$ can be ignored for large $n$). Using the second inequality in Eq.~\eqref{AEP} and $H^{*}=\mathcal{D}_{R}(\psi^{\otimes t})$, it is straightforward to show that
     \begin{equation*}
     \left\vert \frac{\log_{2}\vert \mathcal{A}^{*}_{\epsilon'}\vert}{l} - \mathcal{D}_{R}(\psi^{\otimes t})\right\vert < \epsilon'-\frac{\log_{2} (1-\epsilon')}{l}.
     \end{equation*}
  By dividing the above equation by $t$ and choosing $\epsilon'$ such that $\epsilon'-l^{-1}\log_{2} (1-\epsilon')<\delta ~ t$ (there always exists such $\epsilon'>0$ for any $l,~t,~\delta>0$.), we have
     \begin{equation*}
     \left\vert\frac{\log_{2}\vert \mathcal{A}^{*}_{\epsilon'}\vert}{n} - \frac{\mathcal{D}_{R}(\psi^{\otimes t})}{t}\right\vert<\delta.
     \end{equation*}
  Finally, $\mathcal{D}_{R}(\psi^{\otimes t})~ t^{-1}$ approaches $\mathcal{D}^{\infty}_{R}(\psi)$ as $t\rightarrow\infty$ by the definition of $\mathcal{D}_{R}^{\infty}$, and it completes the proof of the theorem.

 \section{Applications and examples}
  In this section, we present upper and lower bounds for the GHZ entanglement cost (equivalently, the optimal rate) and examine them for several examples. We also compare the LOCC procedure $\mathcal{L}_{D}$ with another procedure that uses singlets as resource. Finally, we discuss the application of $\mathcal{L}_{D}$ to general mixed states.
 \subsection{Upper and lower bounds for the GHZ entanglement cost}
 We denote by $E_{c}$ the GHZ entanglement cost. The existence of $\mathcal{L}_{D}$ implies that $\mathcal{D}_{R}^{\infty}$ is an upper bound for $E_{c}$. In addition, one can easily show that the regularized version of any entanglement measure $\mathcal{E}$ that is non-increasing under LOCC is a lower bound for $E_{c}$, provided that  $\mathcal{E}^{\infty}(GHZ)=1$ and it satisfies the continuity condition, i.e., if $F(\rho^{\otimes n}, \sigma ^{\otimes n})\rightarrow 1$ as $n\rightarrow \infty$, then $\frac{1}{n}\vert \mathcal{E}(\rho^{\otimes n})-\mathcal{E}(\sigma^{n})\vert\rightarrow 0$ as $n\rightarrow\infty$. The relative entropy of entanglement defined in Eq.~\eqref{entanglement} satisfies the continuity condition~\cite{Donald99} and $\mathcal{E}_{R}^{\infty}(GHZ)=1$ that leads to
 \begin{align*}
 \mathcal{E}^{\infty}_{R}(\psi) \leqslant E_{c}(\psi)\leqslant \mathcal{D}_{R}^{\infty}(\psi).
 \end{align*}
  For generalized GHZ states, $\sqrt{p}\vert 000\rangle+\sqrt{1-p}\vert 111\rangle$, the two bounds coincide, so that our procedure is optimal and $E_{c}=\mathcal{D}^\infty_{R}=-p\log_{2} p-(1-p)\log_{2}(1-p)$. In addition, it is known that GHZ states can be distilled from the generalized GHZ states at the same rate~\cite{Bennet00}.

  The GHZ entanglement cost can be either greater or less than 1 as $\mathcal{E}^{\infty}_{R}$ and $\mathcal{D}^{\infty}_{R}$ can be so.  For instance, the GHZ entanglement cost of generalized GHZ states is less than or equal to one as we have shown. For the state $(\vert000\rangle+\vert+11\rangle)/\sqrt{2}$ where $\vert+\rangle=(\vert0\rangle+\vert1\rangle)/\sqrt{2}$, we have $\mathcal{E}^{\infty}_{R}=1$ from $\mathcal{E}^{\infty}_{A:B:C}\geqslant \mathcal{E}^{\infty}_{AB:C}+\mathcal{E}^{\infty}_{A:B}$ (see Ref.~\cite{Yang13,Plenio01} for the inequality). In addition, one can easily see that a single copy of the GHZ state can be converted to a single copy of the state $(\vert000\rangle+\vert+11\rangle)/\sqrt{2}$ by LOCC, so $E_{c}=1$. The regularized discord $\mathcal{D}^{\infty}_{R}$ is not known, but $\mathcal{D}_{R}=1.5$~\cite{Yang13}, so that $E_{c}=1\leqslant \mathcal{D}_{R}^{\infty} \leqslant 1.5$.
  In the case of the W state, $\vert W\rangle=(\vert 001\rangle+\vert 010\rangle +\vert 100\rangle)/\sqrt{3}$, $\mathcal{E}^{\infty}_{R}=2\log_{2}3-2\approx 1.170$~\cite{Zhu10} and $\mathcal{D}_{R}=\log_{2}3\approx 1.585$, so that $1.170<E_{C}<1.585$.

  \subsection{Comparison with preparation of states from singlets} For comparison, we consider a LOCC procedure that prepares a multipartite pure state $\psi^{\otimes n}_{\mathbf{Q}}$ from the singlets shared among $\mathbf{Q}$~\cite{Galvao00}. In this procedure, a single party, say the first party, locally prepares the pure state $\psi^{\otimes n}_{\mathbf{Q}}$ and uses quantum teleportation~\cite{Bennet93} and data compression~\cite{Schumacher95} to distribute the state. For the tripartite case, it requires $nS_{2}$ singlets between $Q_{1}$-$Q_{2}$ and $nS_{3}$ singlets between $Q_{1}$-$Q_{3}$, where $S_{i}$ denotes the von Neumann entropy of the $i$-th subsystem. Therefore, it requires $n(S_{2}+S_{3})$ singlets in total.
  Considering all the permutations of $Q_{1}$, $Q_{2}$ and $Q_{3}$, the consumption rate of singlets is
     \begin{equation*}
     R_{T}=S_{1}+S_{2}+S_{2}-\max \{S_{1}, S_{2}, S_{3}\}.
     \end{equation*}
Since a singlet is obtainable from a single copy of a GHZ state, this procedure can also be achieved by consuming GHZ entanglement at rate $R_{T}$. We have no proof for $\mathcal{D}^{\infty}_{R}\leqslant R_{T}$ for general states. However, it has been shown in Ref.~\cite{Yang13} that $\mathcal{D}_{R}\leqslant R_{T}$ (so $\mathcal{D}^{\infty}_{R}\leqslant R_{T}$) for a few kinds of three-qubit states including the generalized W states $\alpha\vert001\rangle+\beta\vert010\rangle+\gamma\vert100\rangle$, and generalized GHZ states $\alpha\vert000\rangle+\beta\vert111\rangle$. For instance, for the W state, $\mathcal{D}_{R}\approx 1.585$ and $R_{T}\approx 1.837$. We also realize that $\mathcal{D}^{\infty}_{R}=\frac{1}{k-1}R_{T}$ for the $k$-qubit generalized GHZ states, $\alpha\vert 0^{\otimes k}\rangle+\beta\vert 1^{\otimes k}\rangle$, where we have used $R_{T}=\big[\sum_{i=1}^{k} S_{i}\big]-\max\{S_{1},\dotsc,S_{k}\}$. In the case of the $k$-qubit W state $(\vert 00\dotsb1\rangle+\vert 00\dotsb10\rangle+  \dotsb+\vert100\dotsb0\rangle)/{\sqrt{k}}$, one can check that $\mathcal{D}_{R}\leqslant R_{T}$ for any $k$.

\subsection{Application of $\mathcal{L}_{D}$ to general mixed states}
  Before presenting how to prepare general mixed states by using $\mathcal{L}_{D}$, we note that the relation $E_{C} \leqslant  \mathcal{D}_{R}^{\infty}$ does not hold for general mixed states. This can be shown by taking a counter example of a bipartite state,
 \begin{equation*}
 (1-2p)\vert \Phi^{+}\rangle \langle \Phi^{+} \vert +p\big(\vert 00\rangle\langle 00\vert +\vert11\rangle\langle11\vert\big)
 \end{equation*}
 with $\vert \Phi^{+}\rangle = \sqrt{1/2}(\vert 00\rangle +\vert 11\rangle)$ and $0\leqslant p\leqslant1/2$. The entanglement cost of the state is  $E_{C}=H_{2}(1/2+\sqrt{p(1-p)})$~\cite{Vidal02} where $H_{2}(x)=-x\log_{2}x-(1-x)\log_{2}(1-x)$, and the relative entropy of discord is $\mathcal{D}^{\infty}_{R}=1-H_{2}(p)$. One can find that $E_{C}>\mathcal{D}^{\infty}_{R}$ for any $0<p<1/2$.

For a general mixed state $\sigma$, our procedure can be applied to prepare the pure states $\psi_{i}$'s such that $\sigma=\sum_{i} p_{i}\psi_{i}$. Then, the mixed state $\sigma$ can be obtained by classically mixing them. This technique has already been used to generalize the entanglement cost of bipartite pure states to bipartite mixed states~\cite{Horodecki09}. In our case, the rate of the GHZ entanglement consumption is given by
 \begin{equation}
 R_{D}(\rho)=\lim_{t\rightarrow \infty} \frac{r_{D}(\rho^{\otimes t})}{t},
 \end{equation}
 where
 \begin{equation}
r_{D}(\sigma)=\inf \left\{ \sum_{i}p_{i}  \mathcal{D}_{R}(\psi_{i}): \sigma=\sum_{i} p_{i} \psi_{i}\right\}.
\label{eq:rd}
 \end{equation}
  We see that the rate $R_{D}$ vanishes for fully separable states. For bipartite systems, discord $\mathcal{D}_{R}(\psi_{i})$  in Eq.~(\ref{eq:rd}) is equal to the entropy of entanglement, so $R_{D}(\rho)$ is reduced to the regularized version of the entanglement of formation~\cite{Bennet96-2}. It is known that the regularized entanglement of formation is equal to the entanglement cost for bipartite states~\cite{Hayden01}.

 \section{Conclusion}
We have suggested a LOCC procedure that asymptotically prepares an arbitrary pure state from GHZ states where the conversion rate is found  to be the multipartite quantum discord.
It reveals that the GHZ entanglement cost of preparing a multipartite pure state is not higher than a multipartite quantum discord of the state.
Our work provides an operational interpretation of multipartite quantum discord in relation to a multipartite entanglement cost.

\section*{acknowledgements}
 The authors thank Dr.~Kimin Park, Dr.~Animesh Datta and Dr.~Wonmin Son for useful discussions. This work was supported by the National Research Foundation of Korea (NRF) grant funded by the Korea government (MSIP) (No. 2010-0018295).


\begin{thebibliography}{1}
\bibitem{A} As an alternative approach to quantification of multipartite entanglement, one may consdier a minimal reversible entanglement generating set~\cite{Bennet00}, which is a set of a minimum number of states allowing asymtptotocally reversible conversion to any multipartite pure state. However, it left as an open problem to find a such set.
\end{thebibliography}
\end{document}